\documentclass[aps,superscriptaddress,twocolumn,twoside,floatfix,pra,nofootinbib,a4paper]{revtex4-1}
\usepackage{times}
\usepackage{epsfig}
\usepackage{amsfonts}
\usepackage{amsmath}
\usepackage{amssymb}
\usepackage{amsthm}
\usepackage{color}
\usepackage{multirow}
\usepackage[normalem]{ulem}
\newcommand{\stkout}[1]{\ifmmode\text{\sout{\ensuremath{#1}}}\else\sout{#1}\fi}
\usepackage{latexsym}
\usepackage{mathrsfs}
\usepackage{natbib}
\usepackage{verbatim}
\usepackage[T1]{fontenc}
\usepackage{float}
\usepackage{subfigure}

\usepackage{graphicx}
\usepackage{xcolor}

\usepackage[colorlinks=true,linkcolor=blue,citecolor=magenta,urlcolor=blue]{hyperref}

\DeclareMathOperator{\Tr}{tr}

\newcommand{\ket}[1]{|#1\rangle}

\newcommand{\braket}[2]{\langle#1|#2\rangle}
\newcommand{\bracket}[3]{\langle#1|#2|#3\rangle}
\newcommand{\ketbra}[2]{|#1\rangle\langle#2|}

\begin{document}


\title{Semi-device-independent framework based on restricted distrust in prepare-and-measure experiments}

\author{Armin Tavakoli}
\email{armin.tavakoli@oeaw.ac.at}
\affiliation{Institute for Quantum Optics and Quantum Information - IQOQI Vienna, Austrian Academy of Sciences, Boltzmanngasse 3, 1090 Vienna, Austria}

\begin{abstract}
A semi-device-independent framework for prepare-and-measure experiments is introduced in which an experimenter can tune the degree of distrust in the performance of the quantum devices. In this framework, a receiver operates an uncharacterised measurement device and a sender operates a preparation device that emits states with a bounded fidelity with respect to a set of target states. No assumption on Hilbert space dimension is required. The set of quantum correlations is investigated and bounded from both the interior and the exterior. Furthermore, the optimal performance of  quantum state discrimination with bounded distrust is derived and applied to certification of detection efficiency. Quantum-over-classical advantages are demonstrated and the magnitude of distrust compatible with such advantages is explored. Finally, efficient schemes for semi-device-independent random number generation are developed.
\end{abstract}

\date{\today}
\maketitle

\textit{Introduction.---}
Quantum information protocols often assume the precise control of quantum devices. Precise control is, however, an idealisation that experiments only can aspire to approximate. In contrast, the device-independent  (DI) approach to quantum information processing uses violations of Bell inequalities to perform quantum information protocols without requiring any characterisation of the involved quantum devices. Nevertheless, this stringent approach faces substantial experimental obstacles due to its demanding requirements.

Partly motivated by finding a compromise between the black-box spirit of device-independence and the experimental advantages of conventional protocols, and partly motivated by understanding quantum communications, much research attention has been directed at semi-DI quantum information processing. The semi-DI approach is commonly investigated in simple prepare-and-measure experiments in which a sender prepares states and a receiver measures them. Up to a weak and reasonable assumption, the resulting correlations are analysed without requiring any detailed characterisation of the involved devices. 

The most thoroughly researched semi-DI setting is that in which only the Hilbert space dimension is known. This has led to protocols for quantum key distribution \cite{Pawlowski2011, Woodhead2015}, random number generation \cite{Li2011, Li2012}, random access coding \cite{Ambainis1998, Ambainis2008, Tavakoli2015}, numerous  quantum certification protocols \cite{Tavakoli2018, Farkas2019, FakeTriangle, Mohan2019, Miklin2020b, Mironowicz2019, Spee2020, TavakoliSingle, Miklin2020, Carmeli2020, Tavakolieaaw6664} and several experiments \cite{Lunghi2015, Mironowicz2016, Martinez2018, Tavakolieaaw6664, Foletto2020, Anwer2020, Farkas2020}. More recently, also alternative settings have been investigated, based on a bound on the overlap \cite{Brask2017, Wang2019, Shi2019}, energy \cite{VanHimbeeck2017, VanHimbeeck2019, Rusca2019, Tebyanian2020} and information content \cite{Tavakoli2020, Tavakoli2020info} of the states. 

Here, we introduce a framework for semi-DI quantum information processing in which the only assumption is based on the experimenter estimating a bound on how accurately the prepared states correspond to the ideal states targeted in the lab. We model this through a bound on the fidelity between the lab states and the target states. Thus, this tunable distrust (or lack of control) corresponds to a physically observable quality estimate of the preparation procedure.

In what follows, we introduce the framework and proceed to analyse quantum correlations under bounded distrust. We show that the set of correlations can be efficiently bounded from both the interior and the exterior via semidefinite programming (SDP) methods. Then, focusing on quantum state discrimination, which is the simplest scenario, we analytically determine the optimal performance under bounded distrust. This is applied to construct experimentally friendly semi-DI certification of the detection efficiency of a memoryless measurement device. Next, we investigate hybrid models based on classical measurement devices and show that quantum correlations can elude such models even at substantial degrees of distrust. Moreover, we investigate semi-DI random number generation and show that high rates of randomness can be obtained at experimentally realistic levels of distrust. 

\textit{Framework.---}
Consider a scenario in which Alice and Bob independently select inputs $x$ and $y$ respectively. Alice prepares a quantum state $\rho_x$  which she sends to Bob who performs a measurement $\{M_{b|y}\}$ with outcome $b$. When the experiment is repeated in many independent rounds, the correlations are described by the probability distribution $p(b|x,y)=\Tr\left(\rho_x M_{b|y}\right)$. We may also permit the devices of Alice and Bob to be classically correlated through a shared parameter $\lambda$. This leads to the more general probability distribution 
\begin{equation}\label{born}
p(b|x,y)=\sum_\lambda p(\lambda)\Tr\left(\rho_x^{(\lambda)} M_{b|y}^{(\lambda)}\right).
\end{equation}

Suppose that Alice's intention is to prepare a particular set of target states $\{\ket{\psi_x}\}$.  However, her preparation device is subject to a degree of imperfection due to the lack of flawless control. Moreover, her device, or the states it emits (before reaching Bob), could be maliciously influenced. We quantify the accuracy of the preparation procedure through the fidelities 
\begin{equation}\label{fid}
F_{\psi_x}\equiv \bracket{\psi_x}{\rho_x}{\psi_x},
\end{equation} 
where $\rho_x=\sum_\lambda p(\lambda)\rho_x^{(\lambda)}$ is the average state and the target states $\ket{\psi_x}$ are embedded into the arbitrary, but finite, dimensional Hilbert space of $\rho_x$. An ideal procedure ($\rho_x=\ketbra{\psi_x}{\psi_x}$) is represented by $F_{\psi_x}=1$. In contrast, a smaller fidelity signifies that the lab states deviate further from the target states. In our model, we let the experimenter provide a bound on the degree of distrust in their preparation procedure. Specifically, we consider that the fidelities are subject to a lower bound of the form
\begin{equation}\label{assump}
F_{\psi_x}\geq 1-\epsilon_x,
\end{equation}
where $\epsilon_x\in[0,1]$ is the distrust in each state preparation. The fidelity \eqref{fid} can then be interpreted as the probability of obtaining the first outcome of the measurement $\{\ketbra{\psi_x}{\psi_x},\openone-\ketbra{\psi_x}{\psi_x}\}$ when performed on $\rho_x$. Therefore, if there are no side-channels used to send information about $x$, the choice of $\epsilon_x$ can based on direct observation.

\textit{Quantum correlations.---} We develop tools to analyse quantum correlations under bounded distrust. This analysis is considerably simplified by first identifying three key properties. 
(i) The shared parameter $\lambda$ can be absorbed in the preparations by sending the classical-quantum state  $\rho_x=\sum_\lambda p(\lambda)\rho_x^{(\lambda)}\otimes  \ketbra{\lambda}{\lambda}$. Bob learns $\lambda$ by measuring the second register and then proceeds to apply $\{M_{b|y}^{(\lambda)}\}$ to the first register to generate the correlations \eqref{born}. The fidelity is preserved since $F_{\psi_x}=\Tr\left((\ketbra{\psi_x}{\psi_x}\otimes \openone)\rho_x\right)=\sum_\lambda p(\lambda)\bracket{\psi_x}{\rho_x^{(\lambda)}}{\psi_x}$.
(ii) We can w.l.g restrict to considering only pure states $\rho_x$.  Due to Uhlmann's theorem \cite{Uhlmann1976}, for every mixed $\rho_x$ there exists a pure state  $\ket{\phi_x}$  such that the fidelity \eqref{fid} is preserved, i.e.~$F_{\psi_x}=|\braket{\psi_x}{\phi_x}|^2$. Then, by considering measurements that act trivially on the ancillary space of the purification, we recover the quantum correlations; $\Tr\left(\rho_xM_{b|y}\right)=\bracket{\phi_x}{M_{b|y}\otimes \openone}{\phi_x}$. (iii) We can w.l.g restrict to considering only states of dimension $2n$, where $n$ is the number of possible inputs of Alice. This follows from the fact that $m$ pure states span at most an $m$-dimensional subspace of Hilbert space and that our problem involves a total of $2n$ such states. Furthermore, based on systematic numerical evidence discussed in Appendix~\ref{AppDimension}, we conjecture that  the dimension can further be restricted to $n$. 

Equipped with these properties, we investigate linear correlations functions. These are written 
\begin{equation}\label{witness}
\mathcal{W}=\sum_{b,x,y} c_{bxy}p(b|x,y),
\end{equation}
where $c_{bxy}$ are real coefficients. For a general function $\mathcal{W}$, how can we determine the values attainable in quantum theory for a given set of target states $\{\ket{\psi_x}\}$ and a given set of distrust parameters $\{\epsilon_x\}$? We answer this by providing generally applicable methods to establish both lower and upper bounds on the extremal (for simplicity, the maximal) quantum value of the function, which we denote $\mathcal{W}^\text{Q}$.

Any set of states $\{\rho_x\}$ and measurements $\{M_{b|y}\}$ that respect the constraint \eqref{assump} imply a lower bound on $\mathcal{W}^\text{Q}$. Systematic and increasingly accurate lower bounds can be obtained via alternating convex search. This follows from the fact that for fixed measurements, the optimisation problem  $\max_{\{\rho_x\}} \mathcal{W}$ over states $\rho_x$ of dimension $2n$ subject to the constraints \eqref{assump} is an SDP. Similarly, fixing the states to those found optimal by this SDP, the optimisation  $\max_{\{M_{b|y}\}} \mathcal{W}$ over the measurements also consitutes an SDP. Since SDPs can be efficiently evaluated \cite{Vandenberghe1996}, this routine of two SDPs can be iterated to bound $\mathcal{W}^\text{Q}$ from below.

The task of bounding $\mathcal{W}^\text{Q}$ from outside the quantum set is less straightforward. Nevertheless, this can be achieved through a hierarchy of SDP relaxations of the task.  For this purpose, we exploit that we can limit the analys to pure states of dimension $2n$. SDP relaxations of quantum correlations subject only to such dimensional constraints were introduced in Ref.~\cite{Navascues2015, Navascues2015b} based on  randomly sampling from the state and measurement spaces. In Appendix~\ref{AppHierarchy}, we show that this method can be extended to also incorporate the constraints \eqref{assump}. The key adaptation is that the target states are included the sampling procedure in such a way that the fidelities \eqref{fid}, which are merely quantum probabilities, explicitly appear as elements of the final SDP matrix. Notably, this algorithm can be implemented very efficiently by appropriately adapting the methods of Ref.~\cite{Tavakoli2019}. 

Remark: the methods, for bounding the quantum set from the interior and the exterior respectively, have been explored in several case studies and were almost always found to produce coinciding bounds, thus identifying $\mathcal{W}^\text{Q}$ up to solver precision. This attests to their usefulness in practice.

\textit{State discrimination with distrust.---} The simplest task relevant to the distrust-bounded framework is that of quantum state discrimination. Alice has two target states ($x\in\{1,2\}$) which, w.l.g, can be chosen as qubit states with Bloch vectors $\vec{n}_1=(0,0,1)$ and $\vec{n}_2=(\sin\theta,0,\cos\theta)$ for some $\theta\in[0,\pi]$. Bob's aim is to guess their label, i.e.~to output $b=x$. The average success probability is $\mathcal{W}_\text{sd}=\frac{1}{2}p(1|1)+\frac{1}{2}p(2|2)$.
 The well-known textbook scenario corresponds to the special case in which the distrust parameter $\epsilon\equiv \epsilon_1=\epsilon_2=0$, i.e.~Alice's states are known. In contrast, when $\epsilon>0$, Bob attempts discrimination without knowing the precise set of states sent by Alice. 

Correlations in distrust-bounded state discrimination can be analysed by solely analytical means, i.e.~without employing the  above discussed numerical methods. In Appendix~\ref{App212}, the following optimal success probability is derived for any choice of target state ($\theta$) and any distrust parameter ($\epsilon$):
\begin{equation}\label{SD}
\mathcal{W}_\text{sd}^\text{Q}=\frac{1}{2}\left(1+\sin\frac{\theta}{2}\right) +\left[\sqrt{\epsilon(1-\epsilon)}\cos\frac{\theta}{2}-\epsilon\sin\frac{\theta}{2}\right], 
\end{equation}
for $\epsilon\leq \frac{1}{2}\left(1-\sin\frac{\theta}{2}\right)$ and otherwise $\mathcal{W}_\text{sd}^\text{Q}=1$. The expression can be interpreted as an $\epsilon$-dependent correction (second term) to the Helstrom bound  \cite{Helstrom} (first term) which governs conventional state discrimination. Thus, the consequence of not assuming a specific model for how the allowed distrust is manipulated is that the measured correlations become increasingly suboptimal as $\epsilon$ increases. 

%

\textit{Certification of detection efficiency.---} The simplicity of the state discrimination protocol makes it a natural platform for application in semi-DI certification of detection efficiency. Notably, such certification has previously been considered in a dimension-based semi-DI framework \cite{TavakoliSingle}; however the present framework is based on a more natural assumption, experimentally simpler and gives stronger certification.

A simple model of a measurement device endows it with a detection efficiency $\eta\in[0,1]$ which is the probability that it successfully detects an incoming physical system. Consequently, the measurement effectively has three outcomes, $\{\tilde{M}_1,\tilde{M}_2,\tilde{M}_\emptyset\}$, where $\tilde{M}_\emptyset$ represents failed detection. The most general measurement therefore takes the form $M_b=\sum_{\tilde{b}=1,2,\emptyset} p(b|\tilde{b})\tilde{M}_{\tilde{b}}$ for some post-processing $p(b|\tilde{b})$  determining the final outcome $b\in\{1,2\}$. We assume that the efficiency is independent of the incoming state, i.e.~$\forall \rho$: $\Tr(\rho \tilde{M}_\emptyset)=1-\eta$, which implies  $\tilde{M}_\emptyset=\left(1-\eta\right)\openone$.  In state discrimination, it is optimal to map the outcome $\emptyset$ into one of the outcomes $b\in\{1,2\}$, while otherwise setting $b=\tilde{b}$. Therefore, given ($\theta,\epsilon,\eta$), the optimal success probability is  $\eta W^\text{Q}_\text{sd}+\frac{1-\eta}{2}$. Upon observing $\mathcal{W}_\text{sd}$ in the lab, one certifies  that $\eta\geq \frac{2\mathcal{W}_\text{sd}-1}{2\mathcal{W}^\text{Q}_\text{sd}-1}$. 

We investigate the usefulness of this bound by modelling realistic imperfections. Suppose that Alice prepares noisy target states; $\rho_x=v\ketbra{\psi_x}{\psi_x}+\frac{1-v}{2}\openone$ for some visibility $v\in[0,1]$. This corresponds to a distrust of  $\epsilon=\frac{1-v}{2}$. Also, suppose that Bob's optimal measurement, when successful, is perturbed by an alignment error of angle $\delta$ in the Bloch sphere and that the true detection efficiency is $\eta_\text{true}$. Then, the certified bound on the detection efficiency becomes
\begin{equation}\label{etabound}
\eta \geq \begin{cases}
&\frac{v\eta_\text{true} \cos\delta}{v+\sqrt{1-v^2}\cot\frac{\theta}{2}} \quad \text{if } v\geq \sin\frac{\theta}{2} \\
& v\eta_\text{true}\cos\delta\sin\frac{\theta}{2} \quad \text{otherwise},
\end{cases}
\end{equation}
which is linear in $\eta_\text{true}$. For instance, if our target states correspond to $\theta=\frac{5\pi}{6}$ and the imperfections are given by $v=99\%$ ($\epsilon=0.5\%$) and $\delta=1\deg$, we obtain $\eta\geq 0.963\eta_\text{true}$, which is a nearly optimal bound.  In contrast, with an order of magnitude larger imperfections ($v=90\%$, $\delta=10\deg$), the bound remains reasonably good; $\eta\geq 0.855\eta_\text{true}$. Note that \eqref{etabound} becomes stronger for more distinguishable target states.

\textit{Correlations from classical measurements.---} How do we describe classical correlations in the distrust-bounded framework?  The source is inherently quantum since the target states generally do not commute. However, we may consider the situation in which the measurement device is classical, i.e.~all measurements are diagonal in the same basis.  Within the quantum formalism, such measurements are written  $M_{b|y}=\sum_{k}p(b|y,k)\ketbra{e_k}{e_k}$, where $\{\ket{e_k}\}$ is some orthonormal basis of Hilbert space. The correlations then take the form 
\begin{equation}\label{classical}
p(b|x,y)=\sum_{k} p(b|y,k)\bracket{e_k}{\rho_x}{e_k},
\end{equation}
which can be interpreted as a post-processing of the outcome obtained from measuring $\rho_x$ in the basis $\{\ket{e_k}\}$. 

How can we bound any given function \eqref{witness} in such a model? As shown in Appendix~\ref{AppHierarchy}, we can w.l.g restrict to considering only the finitely many deterministic post-processing, $p(b|y,k)\in\{0,1\}$ (again, we may restrict to dimension $2n$). This simplification allows us to bound the maximal value of $\mathcal{W}$ by largely recycling the SDP relaxation method previously discussed for bounding quantum correlations. However, now the SDP relaxation is based only on a single quantum measurement and must be considered separately for every deterministic post-processing (see Appendix~\ref{AppHierarchy}).

Interestingly, there exists a critical value $\epsilon_\text{crit}=\frac{n-1}{n}$ of the distrust parameter $\epsilon\equiv \epsilon_x$, at which Alice can send her input to Bob. Then, Bob can classically generate any distribution $p(b|x,y)$. To show this, consider the worst-case scenario in which all target states are identical, $\ket{\psi_x}=\ket{0}$, and choose the preparations $\ket{\phi_x}=\frac{1}{\sqrt{n}}\sum_{j=0}^{n-1}e^{\frac{2\pi i j(x-1)}{n}}\ket{j}$ corresponding to the Fourier basis of $\mathbb{C}^n$. The fidelities are $|\braket{\psi_x}{\phi_x}|^2=\frac{1}{n}=1-\epsilon_\text{crit}$ and $x$ is recovered by measuring the basis $\{\ket{\phi_x}\}$.

\textit{Quantum advantages.---} Quantum correlations that do not admit the form \eqref{classical} constitute a certificate of the impossibility of viewing Bob's set of measurements as a post-processing of a single measurement. It is therefore evident that if such an advantage exists, it requires at least two measurements. This motivates us to go beyond state discrimination and consider a scenario featuring three preparations and two binary-outcome measurements. We choose the function 
\begin{equation}\label{gallego}
\mathcal{W}_\text{322}=E_{11}+E_{12}+E_{21}-E_{22}-E_{32},
\end{equation}
where $E_{xy}=p(0|x,y)-p(1|x,y)$ and select 
the target states $\{\ket{\psi_1},\ket{\psi_2},\ket{\psi_3}\}$ as qubits forming an isosceles triangle in the $xz$-plane of the Bloch sphere. Their Bloch vectors are $\vec{n}_1=(0,0,1)$, $\vec{n}_2=(1,0,0)$ and $\vec{n}_3=\frac{1}{\sqrt{2}}(-1,0,-1)$. 

\begin{figure}
	\centering
	\includegraphics[width=\columnwidth]{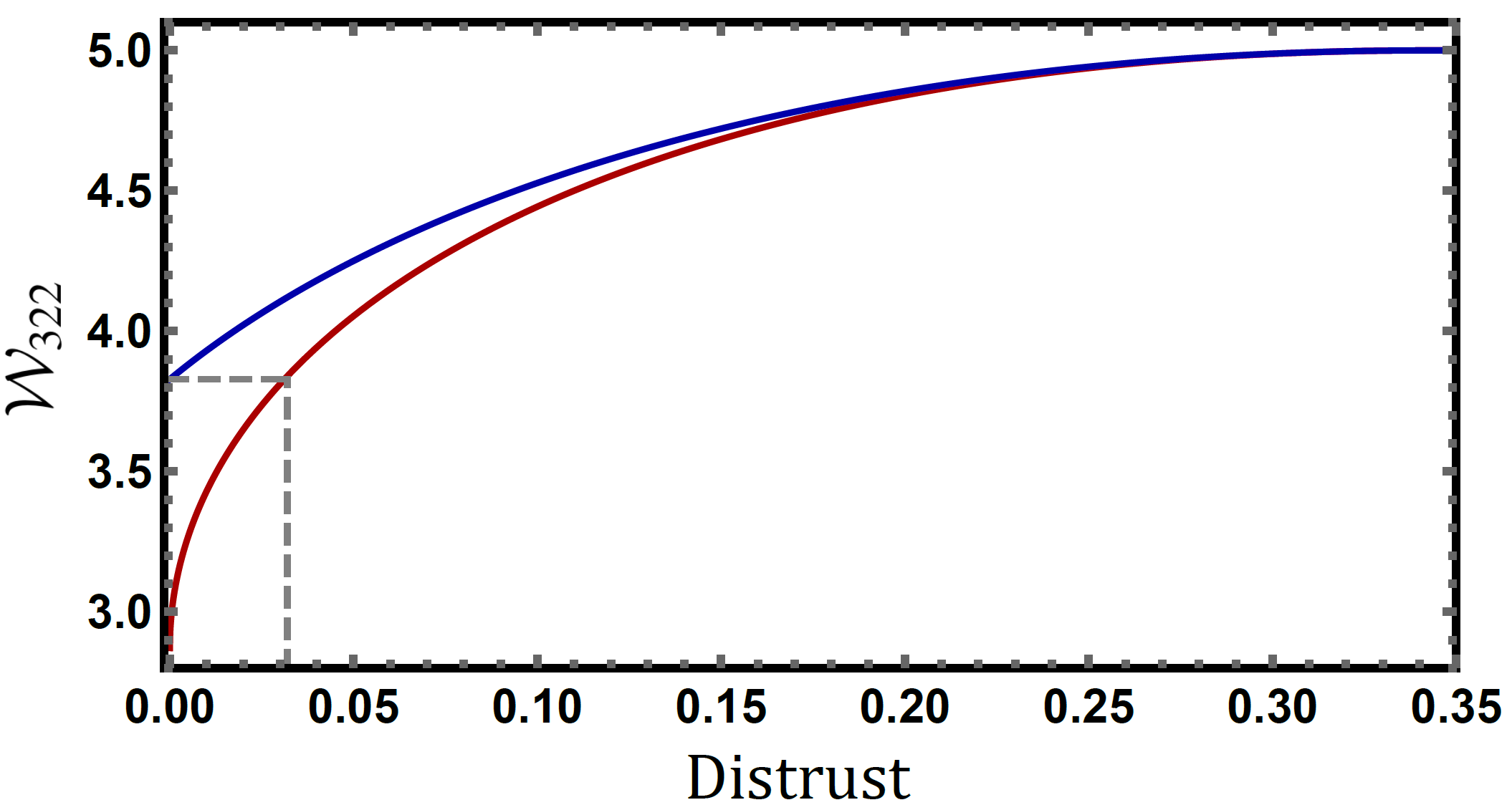}
	\caption{The function $\mathcal{W}_{\text{322}}$ versus the distrust parameter $\epsilon$ for target states forming an isosceles triangle with angles $\{90\deg,135\deg\}$ in the $xz$-plane of the Bloch sphere. The blue curve is the tight quantum bound. The red curve is an upper bound for hybrid quantum-classical models.}\label{Fig_322}
\end{figure}

We have employed alternating convex search together with SDP relaxations of the quantum set of correlations in order to bound $\mathcal{W}_{\text{322}}$ from below and above \cite{Comment} respectively. It is found that these bounds generally coincide, thus constituting a tight bound. Similarly, we have also employed  SDP relaxations to bound the function in models with classical measurements. The results are presented in Figure~\ref{Fig_322}. It is found that quantum theory allows stronger correlations for distrust up  $\epsilon\approx 33\%$. This threshold is marginally lower than the value of $\epsilon$ at which the algebraically maximal value $\mathcal{W}_\text{322}=5$ is attained. Moreover, the maximal quantum value $\mathcal{W}_\text{322}^\text{Q}=1+2\sqrt{2}$, which corresponds to exactly preparing the target states and then performing the optimal measurements, certifies a quantum advantage over classical measurement models even if we supplement the latters with a distrust of up to $\epsilon\approx 3.1\%$. This illustrates a quantum advantage robust to distrusted models with classical measurements. Furthermore, in Appendix~\ref{AppRAC}, we analogously compare correlations based on a Random Access Code and find that the value of $\epsilon$ required to model an ideal quantum protocol with classical measurements is increased to $\epsilon\approx 4.5\%$.

\begin{figure}
	\centering
	\includegraphics[width=\columnwidth]{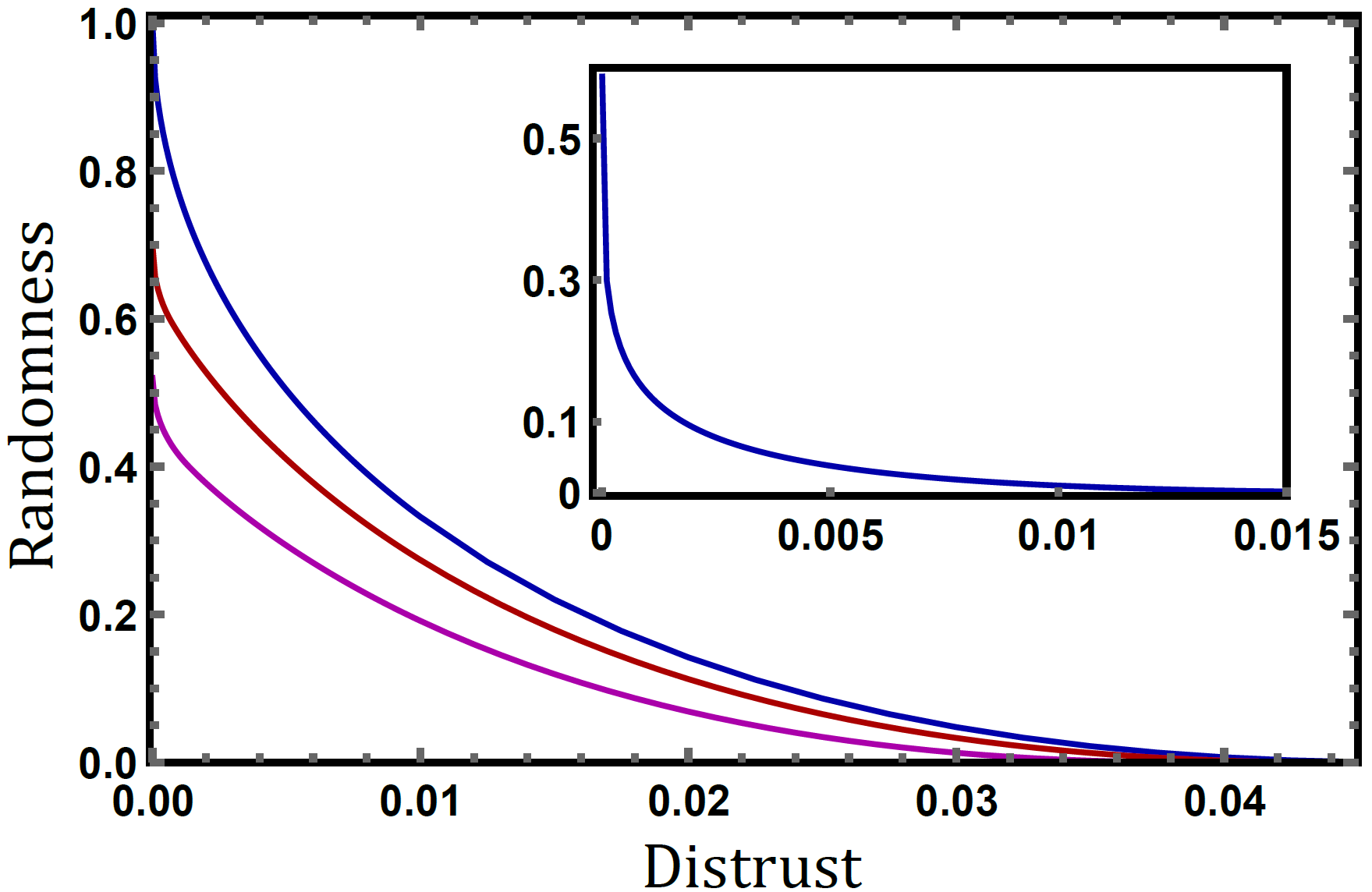}
	\caption{Randomness versus distrust parameter $\epsilon$ for the function $\mathcal{W}_\text{322}$ based on target states forming an isosceles triangle in the $xz$-plane of the Bloch sphere. The randomness is evaluated for the function values $\mathcal{W}_\text{322}^\text{Q}$ (blue), $0.99\mathcal{W}_\text{322}^\text{Q}$ (red) and $0.97\mathcal{W}_\text{322}^\text{Q}$ (purple), where $\mathcal{W}_\text{322}^\text{Q}=1+2\sqrt{2}$ corresponds to the precise target states. Inset plot: randomness versus distrust based on the optimal value of the function $\mathcal{W}_\text{sd}$ for $\theta=\frac{\pi}{5}$.}\label{Fig_randomness}
\end{figure}

\textit{Random number generation.---} We consider the task of semi-DI random number generation in the presence of an external eavesdropper. The eavesdropper can possess a detailed description of the devices on the level of a hidden variable $\lambda$ with distribution $p(\lambda)$. For each $\lambda$, the eavesdropper prepares a quantum realisation for Alice and Bob, thus creating correlations $p_\lambda(b|x,y)$, that on average return Alice's and Bob's observed correlations, $p=\sum_\lambda p(\lambda)p_\lambda$. The task is to certify that Bob's outcome  (for specific inputs $x^*$ and $y^*$) contains some randomness from the eavesdropper's point of view. A standard quantifier of randomness is the conditional min-entropy $H_\text{min}=-\log_2\left(G\right)$, which is based on the eavesdropper's highest probability of guessing Bob's outcome: $G=\max \sum_\lambda p(\lambda)\max_b p_\lambda(b|x^*,y^*)$, where the first maximisation is over $p(\lambda)$ together with all conditional quantum realisations compatible with the observed correlations $p$. If one does not use $p$ but only a linear function $\mathcal{W}[p]$ to certify randomness, then an upper bound on $G$ can be  obtained in the following handy way. Evaluate the $G'(\mathcal{W})=\max \max_{b}p(b|x^*,y^*)$, where the first maximisation now is over all quantum realisations consistent with $\mathcal{W}[p]=\mathcal{W}$. The concave hull of $G'(\mathcal{W})$ is an upper bound on $G$ \cite{Pironio2010, Masanes2011, Nieto2014, Bancal2014}. This method has been used in many previous semi-DI randomness generation protocols  \cite{Li2011,Li2012, Brask2017, Mironowicz2016, Tavakoli2020info}.

In our protocol based on $\mathcal{W}_\text{322}$, using the same target states as earlier, it is favourable to extract randomness from the specific inputs $(x^*,y^*)=(3,2)$. Upper bounds on $G'(\mathcal{W}_\text{322})$ can be obtained systematically using the previously discussed hierarchy of SDP relaxations of the quantum set of correlations. In Figure~\ref{Fig_randomness}, we show the trade-off between the generated randomness and the distrust parameter as obtained from three different function values $\mathcal{W}_\text{322}=k\times (1+2\sqrt{2})$, for $k\in\{0.97, 0.99, 1\}$, where $1+2\sqrt{2}$ is the ideal value obtained in a quantum protocol implementing the precise target states and optimal measurements. We find that even for suboptimal correlations a large amount of randomness can be generated if the distrust is small. Moreover, some randomness is still obtained even when the distrust is around a few percent. In addition, we have illustrated randomness generation based on the simpler state discrimination protocol (see Figure~\ref{Fig_randomness} inset) which is found to give a less robust rate.

The function \eqref{gallego} has been experimentally realised \cite{Ahrens2012} in the context of dimension witnessing using polarisation qubits. Considering the same target states as here, the reported experimental value was $\mathcal{W}_\text{322}^\text{exp}=3.7815\pm 0.0782$. While the distrust could be estimated through explicit additional measurements, let us consider the drastic case in which all the imperfections are attributed to white noise in the preparation device. Then, the average  measured value implies $\epsilon\approx 0.6\%$ from which we can extract $0.052$ bits of randomness. This may be viewed as a proof-of-principle. However, it is relevant to note that recent experiments have performed similar prepare-and-measure experiments achieving visibilities well above $99\%$ in the preparation devices (see e.g.~\cite{Tavakolieaaw6664, Farkas2020}). Such state-of-the-art makes possible substantially higher rates of semi-DI randomness.

\textit{Discussion.---} Here, we have introduced a framework for semi-DI quantum information processing based on a tunable degree of distrust in the quantum devices, investigated the correlations that it may give rise to and harvested these towards quantum information protocols. The tools outlined here are versatile as they apply to general prepare-and-measure scenarios. The introduced framework has two important conceptual features; (i) it tailors the analysis directly to the set of states targeted by the experimenter, and (ii) the notion of distrust is an observable quantity. The first point distinguishes the framework from all previous approaches to semi-DI quantum information. The second point distinguishes it from the standard dimension-based approach \cite{Pawlowski2011}, the overlap-based approach \cite{Brask2017} and the information-based approach \cite{Tavakoli2020}, but not from the energy-based approach \cite{VanHimbeeck2017}. It is then interesting to note that the energy-based framework emerges as a special instance of the distrust-based framework, corresponding to when all target states are  identical. 

A few natural avenues for further research are mentioned. (i) Exploration of other quantum information protocols, e.g.~in cryptography, that are interesting to consider in the distrust-bounded framework. How robust are common quantum information protocols to uncontrolled imperfections? (ii) Experimental realisation of efficient random number generators based on bounded distrust. To this end, it may be relevant to also consider protocols different from those investigated here. (iii) We discussed hybrid models based on quantum sources and classical measurements. Can one define a natural notion of fully classical models? Is it possible to have a quantum advantage over such models when the latters are permitted any degree of distrust smaller  than the critical limit  $\epsilon=\frac{n-1}{n}$?

\begin{acknowledgments}
The author thanks Jef Pauwels, Erik Woodhead, Stefano Pironio and Nicolas Brunner for feedback. This work was supported by the Swiss National Science Foundation through Early PostDoc Mobility fellowship P2GEP2 194800.
\end{acknowledgments}

\bibliography{bib_fidelity}

\appendix

\section{Evidence for restricting to dimension $n$}\label{AppDimension}
In the main text, it was shown that the set of correlations corresponding to a given set of target states and a given set of distrust parameters can without loss of generality be considered with $n$ pure lab states in dimension $2n$. Here, we present numerical evidence supporting the conjecture that the dimension can be further restricted to $n$.

We have considered the following scenarios: $(2,1,2)$, $(2,1,3)$, $(3,2,2)$, $(4,2,2)$ and $(5,4,2)$ respectively, where $(n,m,k)$ denote the number of inputs of Alice, the number of inputs of Bob and the number of outputs of Bob respectively. For each of these prepare-and-measure scenarios, we have sampled $100$ random cases; each corresponding to a random choice of target states, distrust parameters and linear function. W.l.g,  the target states are sampled in $\mathbb{C}^n$. Then, for each case we have numerically searched for the largest function value when Alice's states are restricted to dimension $n$ and when they are restricted to dimension $2n$ respectively. The optimisation is based on the alternating convex search method discussed in the main text and iteration between optimisations over states and measurements is continued until the improvement in the function value, as compared to the  previous round of iteration, is less than $10^{-7}$. It is found that in all five scenarios, each of the $100$ random cases return the same result when considered in dimension $n$ as when considered in dimension $2n$. The largest found discrepancy in favour of using dimension $2n$ found was roughly $10^{-5}$ (which is much smaller than the typical value of the coefficients $c_{bxy}$, each sampled from the interval $[0,1]$) and is reasonably attributed to the precision of the procedure.

\section{Semidefinite relaxations of distrust-bounded correlations}\label{AppHierarchy}
We outline a method to bound the set of quantum correlations subject to bounded distrust. As pointed out in the main text, the set of quantum correlations can, w.l.g, be analysed solely from considering pure quantum states of at most dimension $2n$. This simplification allows us to leverage an already established method for constructing SDP relaxations of the set of quantum correlations subject to a dimensional restriction \cite{Navascues2015}. Here, we show how this method can be adapted to incorporate also the fidelity constraints. Notice that all correlations are trivialised, in the sense that every $p(b|x,y)$ is attainable, if we do not impose the fidelity constraint.

We choose some  function $\mathcal{W}$ on the form given in the main text, a set of target states $\{\ket{\psi_x}\}$ (embedded as rays in $\mathbb{C}^n$) and a set of distrust parameters $\{\epsilon_x\}$. We are interested in addressing the following optimisation problem.
\begin{align}\label{opt}\nonumber
& \qquad \qquad  \qquad\mathcal{W}^\text{Q}=\max_{\{\phi_x\}, \{M_{b|y}\}} \mathcal{W} \\\nonumber
&\text{such that} \quad \ket{\phi_x}\in \mathcal{R}(\mathbb{C}^{2n}), \quad |\braket{\phi_x}{\psi_x}|^2\geq 1-\epsilon_x,\\
& \quad M_{b|y}\geq 0, \hspace{1mm} \text{and}\quad M_{b|y}M_{b'|y}=\delta_{b,b'}M_{b|y},
\end{align} 
where $\mathcal{R}(\mathbb{C}^{2n})$ is the set of rays (normalised vectors) in $\mathbb{C}^{2n}$. Notably, we have restricted the analysis to projective quantum measurements of dimension $2n$. However, positive operator-valued measures can be considered by increasing the dimension to $(2n)^2$, in which every generalised measurement of dimension $2n$ can be written as a projective measurement. Then, we again encounter a problem analogous to \eqref{opt}.

\subsection{Quantum models}

In order to bound $\mathcal{W}^\text{Q}$ from above, we consider the following SDP relaxation.  Define a list of operators $S=\{\openone,\phi_1,\ldots,\phi_n,\psi_1,\ldots,\psi_n,M_{1|1},\ldots,M_{k|m}\}$, where $m$ and $k$ are the number of inputs and outputs for Bob. Also, define a list of monomials $\mathcal{S}$ whose elements are products of the operators appearing in $S$. While $\mathcal{S}$ should, at the very least, include all operators in $S$ (products of length one), the higher-order products that may be included constitute a degree of freedom. From the list of monomials, define a moment matrix $\Gamma_{ij}=\Tr\left(\mathcal{S}_i\mathcal{S}_j^\dagger\right)$. Clearly, this matrix is positive semidefinite and includes the quantum probabilities $p(b|x,y)=\bracket{\phi_x}{M_{b|y}}{\phi_x}$ among its elements. We denote these elements $\Gamma_{bxy}$. Thus, the existence of $\Gamma\geq 0$ is a necessary condition for a given probability distribution to admit a quantum model. Furthermore, since any function $\mathcal{W}$ is a linear combination of probabilities, it can be expressed, on the level of the moment matrix, as a linear combination of the elements
\begin{equation}\label{awit}
\mathcal{W}=\sum_{b,x,y}c_{bxy} \Gamma_{bxy}.
\end{equation}
Moreover, the moment matrix also features the fidelities $|\braket{\phi_x}{\psi_x}|^2=\Tr\left(\phi_x \psi_x\right)$ as explicit elements. These elements are denoted $\Gamma_{x}$. Thus, the bounded distrust can be enforced as a set of linear constraints 
\begin{equation}\label{afid}
\Gamma_x \geq 1-\epsilon_x.
\end{equation}

So far, we have not imposed any structure from the $2n$-dimensional Hilbert space to which our problem \eqref{opt} is constrained. This is achieved using the random sampling method of Ref.~\cite{Navascues2015}.  Here, we summarise this method while referring the reader to Refs~\cite{Navascues2015, Navascues2015b} for further details.

Sample a random set of $2n$-dimensional pure states $(\rho_1,\ldots,\rho_n)$ and a random set of $2n$-dimensional projective measurements $(M_{1|1},\ldots,M_{k|m})$ of a fixed set of ranks. From these, we obtain a sampled operator list $S$ (in which the operators $\openone$ and $(\psi_1,\ldots,\psi_n)$ remain fixed at all times). From the sampled operator list, we compute the associated sampled monomial list $\mathcal{S}$ and use it to compute the sampled moment matrix $\Gamma^{(1)}$. This procedure is repeated many times, leading to a list of independently sampled moment matrices $\{\Gamma^{(1)},\ldots, \Gamma^{(r)}\}$. The sampling process is terminated when the next sampled moment matrix, $\Gamma^{(r+1)}$, is found to be linearly dependent on the set $\{\Gamma^{(1)},\ldots, \Gamma^{(r)}\}$. Then, $\{\Gamma^{(1)},\ldots, \Gamma^{(r)}\}$ constitutes a basis of the space of relevant moment matrices. Thus, the final moment matrix is an affine combination (to preserve normalisation) of the sampled basis:
\begin{equation}\label{moment}
\Gamma=\sum_{i=1}^r s_i \Gamma^{(i)} \quad \text{such that}  \quad \sum_{i=1}^r s_i=1.
\end{equation}
The SDP relaxation of \eqref{opt} amounts to finding the largest value of the function \eqref{awit} under the fidelity constraints \eqref{afid} for a moment matrix \eqref{moment} such that $\Gamma \geq 0$. Conversely, if a value of $\mathcal{W}$ cannot satisfy these constraints then it cannot admit a quantum model. Notably, by increasing the operator products included in the monomial list $\mathcal{S}$, we obtain increasingly strong necessary conditions for a quantum realisation. We remark that the problem has to be solved separately for each of the rank combinations associated to the measurement operators and the largest value among these consitutes the final upper bound on \eqref{opt}.

\subsection{Classical measurement models}
Let us also discuss how the above method applies to bounding correlations in hybrid models, i.e.~models in which the preparation device is quantum and the measurement device is classical. As stated in the main text, correlations in a semi-classical model read $p(b|x,y)=\sum_{k} p(b|y,k)\bracket{e_k}{\rho_x}{e_k}$ for some orthonormal basis $\{\ket{e_k}\}$. Thus, we can express any given linear function as
\begin{equation}
\mathcal{W}= \sum_{x,y,b}c_{bxy}\sum_{k}p(b|y,k)\bracket{e_k}{\rho_x}{e_k}.
\end{equation}
We may decompose the post-processing as a convex combination of determininstic post-processings $p(b|y,k)=\sum_\mu p(\mu) \delta(f_\mu(y,k)=b)$, where $\{f_1,\ldots,f_{k^{2nm}}\}$ is the list of all functions that map the pair $(y,k)$ to the outcome $b$ and $\delta(A=B)$ equals one if $A=B$ and zero otherwise. Thus, we may write our function as
\begin{equation}
\mathcal{W}=\sum_\mu p(\mu) \left[\sum_{x,y,b}c_{bxy}\sum_{k} \delta(f_\mu(y,k)=b)\bracket{e_k}{\rho_x}{e_k}\right].
\end{equation}
This can now be interpreted as a convex combination in $\mu$ of the bracket-expression. Consequently, the optimal value is necessarily attained by some deterministic distribution $p(\mu)\in\{0,1\}$. It is therefore sufficient to consider only deterministic post-processings $p(b|y,k)$. This gives
\begin{equation}
\mathcal{W}=\max_{\mu\in\{1,\ldots,k^{2nm}\} }\left[\sum_{x,y,b}c_{bxy}\sum_{k} \delta(f_\mu(y,k)=b)\bracket{e_k}{\rho_x}{e_k}\right].
\end{equation}
For any given $\mu$, the bracket quantity can be bounded from above using the outlined SDP relaxation method. This entails replacing the operators $(M_{1|1},\ldots, M_{k|m})$, encountered in the quantum case, with a single quantum measurement $\{M_1,\ldots,M_{2n}\}$. The final bound on the function is obtained from selecting the largest bound obtained on the above bracket expression among all values of $\mu$.

\section{Optimal  state discrimination}\label{App212}
\begin{figure}
	\centering
	\includegraphics[width=\columnwidth]{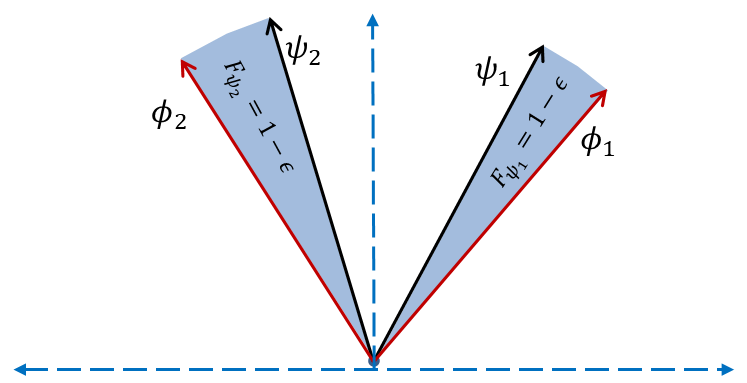}
	\caption{Optimal configuration of vectors for state discrimination with $\epsilon$ distrust. The optimal states $\ket{\phi_1}$ and $\ket{\phi_2}$ are chosen with an angle to the target states $\ket{\psi_1}$ and $\ket{\psi_2}$ respectively, consistent with the allowed distrust. The orientation of the lab states is such that their modulus inner product is minimised.}  \label{Fig_vectors}                  
\end{figure}

We present a derivation of the optimal correlations in quantum minimal error state discrimination for a general pair of target states and any distrust parameter $\epsilon\equiv \epsilon_1=\epsilon_2$. W.l.g, we can choose the first target state as $\ket{0}$, corresponding to a Bloch vector $\vec{n}_1=(0,0,1)$, and the second target state corresponding to a Bloch vector $\vec{n}_2=(\sin\theta,0,\cos\theta)$. Due to the fact that the measurement only has binary outcomes, we can analytically eliminate the measurement from the optimisation:
\begin{multline}
\max_{\rho_1,\rho_2,M_1,M_2} \mathcal{W}=\max_{\rho_1,\rho_2,M_1,M_2} \frac{1}{2}\left(1+\Tr\left((\rho_1-\rho_2)M_1\right)\right)\\
=\max_{\rho_1,\rho_2} \frac{1}{2}\left(1+\lambda_+(\rho_1-\rho_2)\right)=\max_{\phi_1,\phi_2} \frac{1}{2}\left(1+\lambda_\text{max}(\phi_1-\phi_2)\right),
\end{multline}
where we first have  optimally chosen $M_1$ as the projector onto the positive eigenspace of $\rho_1-\rho_2$ and then denoted by $\lambda_+(A)$ the sum of all positive eigenvalues of $A$. In the last step, we have used the fact that we may restrict $\{\rho_1,\rho_2\}$ to be pure states $\{\ket{\phi_1},\ket{\phi_2}\}$, and thus only need to consider the maximal eigenvalue $\lambda_\text{max}$. Furthermore, it is straightforwardly shown that for a pair of pure states, one has
\begin{equation}
\lambda_\text{max}(\phi_1-\phi_2)=\sqrt{1-|\braket{\phi_1}{\phi_2}|^2}.
\end{equation}
Thus, our task is equivalent to minimising the fidelity between $\ket{\phi_1}$ and $\ket{\phi_2}$. To this end, we follow our previous conjecture and choose these states in the same  qubit space as the target states\footnote{The optimality of the conjecture is further supported for this particular task using the SDP hierarchy method.} (see Figure~\ref{Fig_vectors}). 
Using this, and by assigning both the pure qubit states with respective Bloch vectors $\vec{m}_x=(\sin \nu_x,0,\cos\nu_x)$, we can write the optimal function value as
\begin{equation}\label{step}
\max_{\nu_1,\nu_2}\mathcal{W} =\frac{1}{2}\left(1+\left|\sin\left(\frac{\nu_1-\nu_2}{2}\right)\right|\right).
\end{equation}
Now, we consider the fidelity constraints $F_{\phi_x}\geq 1-\epsilon$. These two constraints (for $x=1,2$) can be simplified to the form
\begin{align}
& \cos \nu_1\geq 1-2\epsilon\\
& \cos\left(\theta-\nu_2\right)\geq 1-2\epsilon.
\end{align}
These are equivalent to the following constraints on the Bloch sphere angles:
\begin{align}
& -\arccos\left(1-2\epsilon\right)\leq \nu_1\leq \arccos\left(1-2\epsilon\right)\\
& \theta-\arccos\left(1-2\epsilon\right)\leq \nu_2\leq \theta+\arccos\left(1-2\epsilon\right).
\end{align}
In particular these two inequalities enable the configuration $\nu_1-\nu_2=\pi$ when $\pi-\theta\leq 2\arccos\left(1-2\epsilon\right)$, which is equivalent to $\epsilon\geq  \frac{1}{2}\left(1-\sin\frac{\theta}{2}\right)$. Thus, in this domain, one attains the unit value in \eqref{step}. Non-trivial solutions are obtained when $\epsilon\leq \frac{1}{2}\left(1-\sin\frac{\theta}{2}\right)$. Then, an optimal choice is $\nu_1=-\arccos\left(1-2\epsilon\right)$ and $\nu_2=\theta+\arccos\left(1-2\epsilon\right)$. Inserting this into \eqref{step}, some simplifications lead to the final result
\begin{equation}
\mathcal{W}_\text{sd}^\text{Q}= \frac{1}{2}\left(1+\sin\frac{\theta}{2}\right) +\left[\sqrt{\epsilon(1-\epsilon)}\cos\frac{\theta}{2}-\epsilon\sin\frac{\theta}{2}\right].
\end{equation}

\section{Quantum versus classical correlations in a Random Access Code}\label{AppRAC}

We consider a Random Access Code (RAC). In this task, Alice has four possible inputs represented by two bits, $x=x_1x_2\in\{0,1\}^2$, and Bob has a binary input $y\in\{1,2\}$ and a binary outcome $b\in\{0,1\}$. The task is for Bob to guess the $y$'th bit of Alice. The average success probability is given by the function
\begin{equation}
\mathcal{W}_\text{RAC}=\frac{1}{8}\sum_{x,y}p(b=x_y|x,y).
\end{equation}
A standard quantum RAC is performed using four qubit states forming a square in the $xz$-plane of the Bloch sphere. These correspond to the target states
\begin{align}
& \ket{\psi_{00}}=\ket{0},  & \ket{\psi_{11}}=\ket{1},\\
& \ket{\psi_{01}}=\frac{\ket{0}+\ket{1}}{2}, & \ket{\psi_{10}}=\frac{\ket{0}-\ket{1}}{2}. 
\end{align}
Then, the best measurements of Bob correspond to the observables $\frac{\sigma_z\pm \sigma_x}{\sqrt{2}}$, which lead to the function value $\mathcal{W}_\text{RAC}=\frac{1}{2}\left(1+\frac{1}{\sqrt{2}}\right)\approx 0.854$.

\begin{figure}[h!]
	\centering
	\includegraphics[width=\columnwidth]{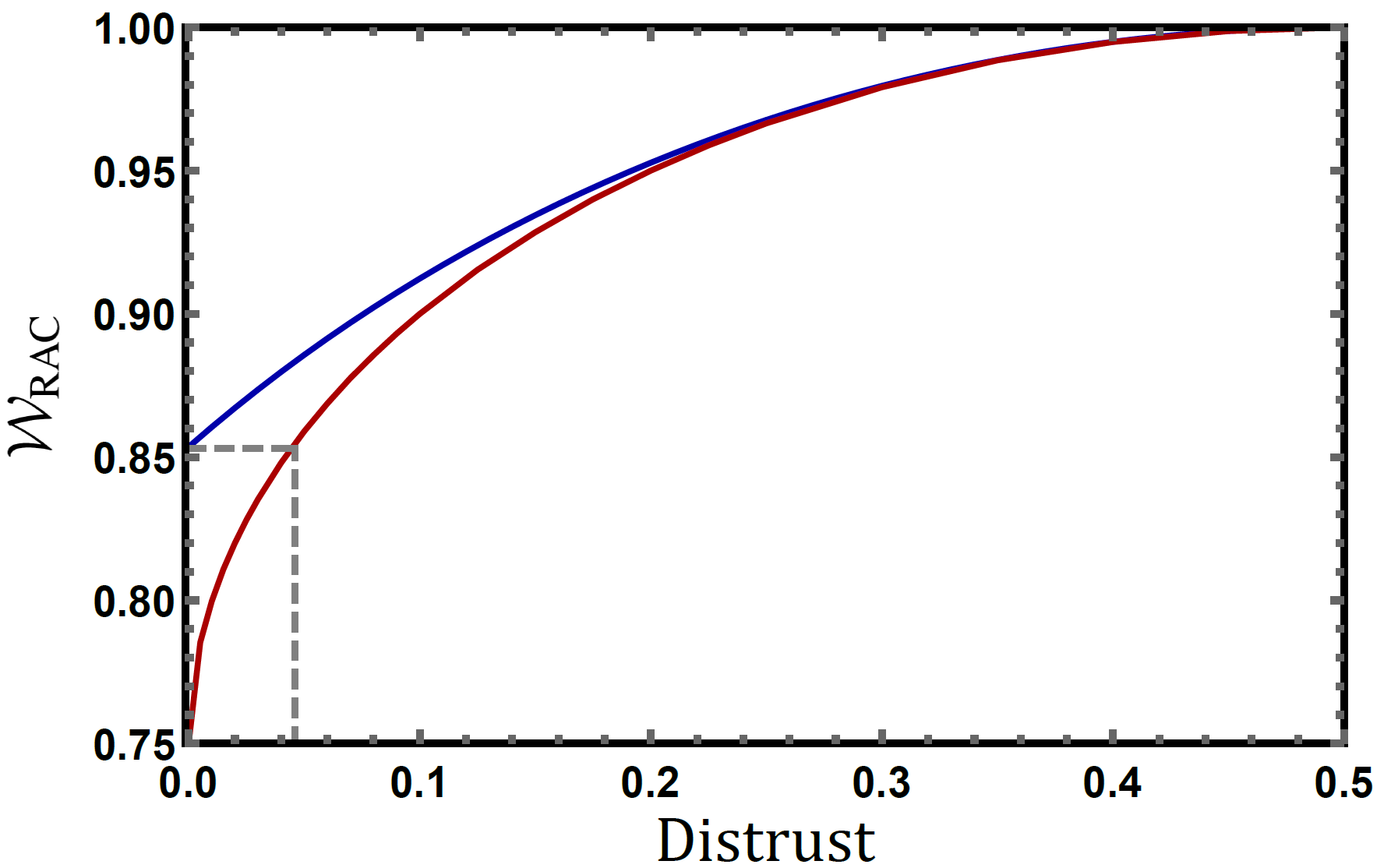}
	\caption{Tight bound on $\mathcal{W}_\text{RAC}$ in quantum models versus the distrust $\epsilon$ (blue). The red curve is an upper bound on the same function in classical models.}  \label{Fig_422}                  
\end{figure}

Following the analysis of the function $\mathcal{W}_\text{322}$ in the main text, we have investigated the maximal value of $\mathcal{W}_\text{RAC}$ attainable in quantum and classical models with a distrust $\epsilon$. The quantum case is addressed using alternating convex search to place lower bounds on $\mathcal{W}_\text{RAC}$ combined with semidefinite relaxations to establish upper bounds. The latter was implemented using the hierarchy level corresponding to the following monomial list $\{\openone,\rho,\psi,M,\rho^2,\rho \psi, \rho M, M^2\}$, which gives an SDP of size $77$. We find that the lower and upper bounds agree up to solver precision, thus precisely characterising $\mathcal{W}_\text{RAC}^\text{Q}$ as a function of $\epsilon$. Similarly, we have used SDP relaxations (with the same level, moment matrix size $61$) to bound the classical function value. The results are illustrated in Figure~\ref{Fig_422}. Both quantum and classical correlations are non-trivial for any $\epsilon<\frac{1}{2}$. Notably, for $\epsilon=\frac{1}{2}$ it is possible for Alice's four states to form an orthonormal basis, thus trivialising the task. Moreover, if we implement the quantum protocol outlined above using the exact target states, then the correlations are found to elude a classical model with a distrust up to $\epsilon\approx 4.5\%$.

\end{document}